\documentclass[sigconf]{acmart}
\usepackage{bm}
\usepackage{algorithm}
\usepackage{algorithmic}
\usepackage[normalem]{ulem}
\useunder{\uline}{\ul}{}
\usepackage{makecell}
\usepackage{graphicx}
\usepackage{float}
\usepackage{subfig}

\AtBeginDocument{%
  \providecommand\BibTeX{{%
    \normalfont B\kern-0.5em{\scshape i\kern-0.25em b}\kern-0.8em\TeX}}}

\copyrightyear{2022}
\acmYear{2022}
\setcopyright{acmcopyright}
\acmConference[CIKM '22] {Proceedings of the 31st ACM International Conference on Information and Knowledge Management}{October 17--21, 2022}{Atlanta, GA, USA.}
\acmBooktitle{Proceedings of the 31st ACM International Conference on Information and Knowledge Management (CIKM '22), October 17--21, 2022, Atlanta, GA, USA}
\acmPrice{15.00}
\acmISBN{978-1-4503-9236-5/22/10}
\acmDOI{10.1145/3511808.3557363}

\acmSubmissionID{0561}

\begin{document}

\title{MetaTrader: An Reinforcement Learning Approach Integrating Diverse Policies for Portfolio Optimization}

\author{Hui Niu}
\authornote{Equal Contribution.}
\email{niuh17@mails.tsinghua.edu.cn}
\affiliation{%
  \institution{Tsinghua University}
  \state{Beijing}
  \country{China}
}

\author{Siyuan Li}
\authornotemark[1]
\email{lisiyuan199511@gmail.com}
\affiliation{%
  \institution{Harbin Institute of Technology}
  \city{Harbin}
  \country{China}}

\author{Jian Li}
\email{lapordge@gmail.com}
\affiliation{%
  \institution{Tsinghua University}
  \state{Beijing}
  \country{China}
}

\begin{abstract}
Portfolio management is a fundamental problem in finance. It involves periodic reallocations of assets to maximize the expected returns within an appropriate level of risk exposure. Deep reinforcement learning (RL) has been considered a promising approach to solving this problem owing to its strong capability in sequential decision making.  However, due to the non-stationary nature of financial markets, applying RL techniques to portfolio optimization remains a challenging problem. Extracting trading knowledge from various expert strategies could be helpful for agents to accommodate the changing markets. In this paper, we propose \textit{MetaTrader}, a novel two-stage RL-based approach for portfolio management, which learns to integrate diverse trading policies to adapt to various market conditions.  In the first stage, MetaTrader incorporates an imitation learning objective into the reinforcement learning framework. Through imitating different expert demonstrations, MetaTrader acquires a set of trading policies with great diversity. In the second stage, MetaTrader learns a meta-policy to recognize the market conditions and decide on the most proper learned policy to follow. We evaluate the proposed approach on three real-world index datasets and compare it to state-of-the-art baselines. The empirical results demonstrate that MetaTrader significantly outperforms those baselines in balancing profits and risks. Furthermore, thorough ablation studies validate the effectiveness of the components in the proposed approach.
\end{abstract}

\ccsdesc[500]{Computing methodologies~Reinforcement learning}
\ccsdesc[500]{Applied computing~Economics}

\keywords{Portfolio Management, Deep Reinforcement Learning, Imitation Learning, Meta-Policy Learning}

\maketitle

\section{Introduction}
\label{sec1}

Portfolio management has long been an important and challenging problem in quantitative trading \cite{markowvitz1952,li2000multiperiod,li2014pmsurvey,ben2021survey}, which aims to maximize the expected returns with acceptable risks through reallocating portfolio weights. Generally, portfolio management is a sequential decision making problem. 

The \textit{Markowitz} model \cite{markowvitz1952}, also called the mean-variance model, is the first theoretical work to formally investigate portfolio optimization. This model formulates portfolio management as a single-period optimization problem, and is then generalized to multi-period optimization in several works \cite{jan1968meanvar, steinbach2001meanvar, li2000multiperiod}.
However, the mathematical solutions for those models require explicit models of both temporal dynamics of individual assets and their co-movements, which are difficult to acquire in practice \cite{ben2021survey}.
Some empirical work devotes to solving the portfolio management problem based on typically observed patterns, including momentum investing \cite{jegadeesh2002csm}, mean reversion \cite{jegadeesh1993blsw}, and multi-factor models \cite{Fama1996multifactor}. Nevertheless, these traditional investment strategies could hardly adapt to the changing markets since they only perform well in certain cases.

More recently, a variety of machine learning methods are developed to address the portfolio management problem. In particular, deep neural networks such as graph models  \cite{cheng2021graphattention,scarselli2008graph,sawhney2021hypergraph} and recurrent networks \cite{hu2017lstm,wang2018lstm} are widely employed to extract temporal and spatial relationships of asset data, benefitting from their strong abilities in representation learning. Several supervised learning methods are proposed to predict price movements based on various kinds of financial information \cite{hu2018news,yoo2021dtml,hou2021contrastive,qin2017darnn}.
Nevertheless, these supervised learning algorithms require data with a set of labels that are associated with certain tasks (e.g. classification and regression).  The design of labels for financial data is nontrivial \cite{de2018advances}, especially in the context of portfolio management. Moreover, as the portfolio weights are generated indirectly, these supervised learning methods suffer from fragile hyper-parameter tuning. 

Efforts have also been made to adopt deep reinforcement learning (RL) techniques to address the portfolio management problem \cite{jiang2017elle, wang2019alphastock, wang2021deeptrader, ding2018ii, shi2021xpm-cikm, huang2022mspm}. Compared to supervised learning, RL is considered a more flexible framework for portfolio management since it takes advantage of various reward functions to achieve better risk-return balancing.
Despite promising results achieved by the previous RL-based methods, applying RL to real-world portfolio management still faces challenges. The non-stationary nature of the financial markets induces a challenging exploration problem for RL algorithms: Agents could hardly fully explore the fluctuate trading environments through maximizing cumulative rewards without additional knowledge, which further influences the performance of the learned policy.

To overcome this challenge, we propose \textit{MetaTrader}, a novel RL-based portfolio management approach incorporating with imitation learning techniques. The proposed approach is inspired by the following facts:
(1) Beyond learning from interactions with the environment, extracting trading knowledge from seasoned experts is also an appealing way to mine profitable patterns in quantitative trading \cite{ding2018ii}. 
(2) Leveraging expert data is a favorable approach to improve the exploration ability of RL algorithms \cite{nair2018overcoming, balaguer2011combining, sun2018truncated, mendonca2019guided}.
(3) Financial companies employ multiple portfolio managers to diversify risks, and  those managers are adept at dealing with  different market conditions \cite{lee2020maps}.
 To summarize, the key insight of MetaTrader is that utilizing knowledge from various trading strategies promotes the agent's adaptability to the changing markets.

Motivated by the above intuitions, MetaTrader decomposes the portfolio optimization process into two phases. The first is a \textit{diverse policy learning} phase, which aims to obtain a set of base policies capable of dealing with various market conditions. To achieve this goal and promote efficient exploration of the complex trading environment, we develop a novel learning objective that combines the RL objective of maximizing expected cumulative rewards with imitating different trading experts. This objective enables learning from interactions with the environments and through extracting knowledge from expert datasets as well.
The second is a \textit{meta-policy learning} phase, where a meta-policy is learned to recognize the market conditions and make decisions on which learned base policy to follow in each holding period. This phase is modeled with the RL framework, and MetaTrader learns to score the base policies obtained in the first phase while freezing their parameters.

The contribution of this paper can be summarized as follows:
\begin{itemize}
    \item To the best of our knowledge, MetaTrader is the first attempt to utilize trading knowledge from \textit{multiple experts} to acquire diverse policies in the portfolio optimization literature. 
    \item MetaTrader learns to select the reasonable base policy to follow by recognizing the market conditions.
 The scores given by the meta-policy reflect its confidence in the profitability of those base policies, which enhances the interpretability of the learned portfolio management decisions. 
    \item Extensive experiments on three well-known stock indexes show that our portfolio management agent learns to adapt to the changing markets properly and achieves superior performance in profitability and risk-return balancing compared to state-of-the-art baselines. Further ablation studies verify the effectiveness of meta-policy learning.
\end{itemize}

This paper is organized as follows: In Section \ref{sec2}, we formulate the portfolio management problem as a Markov decision process and introduce the trading procedure. Afterward, we describe the proposed approach MetaTrader in Section \ref{sec3}, including the detailed objective, structure, and workflow.  Next, in Section \ref{exp}, we conduct extensive experiments to illustrate the effectiveness of MetaTrader. Then, in Section \ref{related}, we review and discuss the related works in portfolio management.
Finally, Section \ref{con} concludes and points out some possible future directions.

\section{Problem Formulation}
\label{sec2}

\begin{figure*}[t!]
  \centering
  \includegraphics[width=0.85\linewidth]{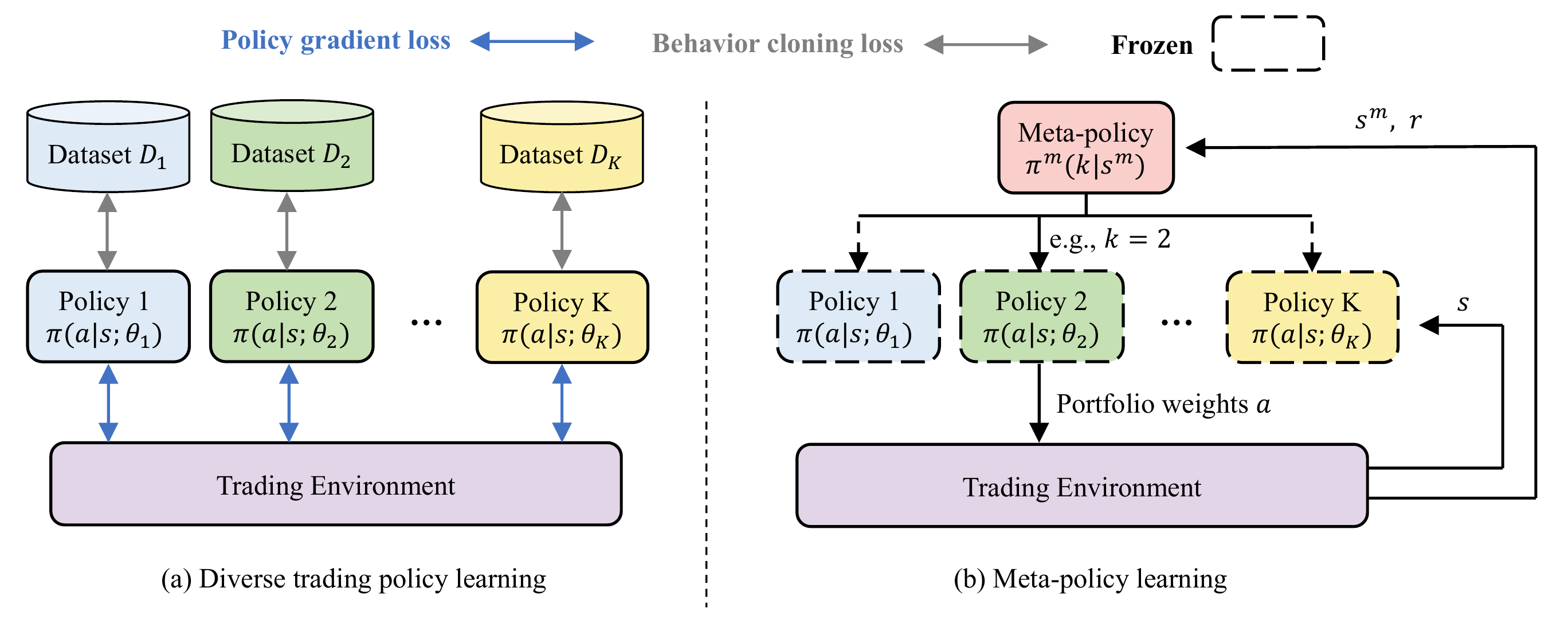}
  \setlength{\belowcaptionskip}{-0.3cm} 
  \caption{The learning framework of MetaTrader.}
  \label{fig1}
\end{figure*}

In this section, we introduce some notations and describe the problem formulation in this work for portfolio management. 

\subsection{Problem Setup}
A financial portfolio is a collection of assets comprised of stocks, bonds, cash, or other forms of assets. 
We consider the investment in $N$ risky assets in a market, e.g., stocks\footnote{For ease of explanation, we may refer to assets as stocks in the following. The proposed approach is applicable to other kinds of risky assets  as well.}.

 Portfolio management is a sequential decision making problem involving periodically reallocating the asset weights.
 Suppose an investor enters the market at time $0$ with an initial wealth $AC_0$, he/she reallocates the asset weights at time $t\times\Delta t$ for $t=0,1, ..., T-1$, where $\Delta t$ is a minimum time unit for investment, for example, one day or one month. The $t$-th holding period lasts from time $t\times\Delta t$ to time $(t+1) \times \Delta t$. 
 The starting time of the $t$-th holding period is called \textit{timestep} $t$, and the ending time of it is called timestep $t+1$. The accumulated capital at timestep $t$ is denoted by $AC_t$. Simiarly, the subscript $t$ denotes timestep $t$ in the following.

This sequential decision making problem is formulated as a Markov decision process (MDP) $M = \langle  \mathcal S, \mathcal A, P, R, \gamma \rangle$, where $\mathcal S$ is the state space, $\mathcal A$ is the $2N$-dimensional action space, $P:\mathcal S \times \mathcal A \times \mathcal S \rightarrow [0, 1]$ is the state transition function specifying the conditional transition probabilities between states, $R: \mathcal S \times \mathcal A \times \mathcal S \rightarrow \mathbb R$ is the bounded reward function and $\gamma \in (0, 1]$ is the discount factor.

The details of the MDP environment are set as follows:
\begin{itemize}
    \item State space $\mathcal S$: A state $s_t \in \mathcal S$ consists of three components $s_t = \{\bm{x}_t^s;\bm{x}_t^m;\bm{x}_t^a\}$, which are generated from different data sources.
    Asset features $\bm{x}_t^s$ describe the information of the assets, including historical asset prices and some classical technical indicators\footnote{The details for the classical technical indicators are specified in the experiment section, Section \ref{exp}.}. 
    Market features $\bm{x}_t^m$ reflect current market conditions using macro indicators such as index trends and market sentiment. 
    Account feature $\bm{x}_t^a$ is the portfolio weight at timestep $t-1$, $\bm{x}_t^a =\bm{a}_{t-1}$. 
    
    \item Action space $\mathcal A$: Both long and short operations can be performed in this formulation. An action $a_t \in \mathcal A$ is the reallocated portfolio weight at timestep $t$. Specifically, ${a}_t =[\bm{w}_t^+;\bm{w}_t^-]^T = [w_{0, t}^+, w_{1, t}^+,$ $ ..., w_{N-1, t}^+; w_{0, t}^-,w_{1, t}^-, ..., w_{N-1, t}^-]^T$, where $w_{i, t}^+ \in [0,1]$ denotes the weight of long position on asset $i$ , $\sum_{i=0}^{N-1}w_{i, t}^+ =1$; and $w_{i, t}^- \in [-1,0]$ denotes the weight of short allocation on asset $i$, $\sum_{i=0}^{N-1}w_{i, t}^-$ $= -\rho_t$.
    Here $\rho_t$ is a learned ratio of the short position. Precisely, at the beginning of the $t$-th holing period,  the total value of the new borrowed stocks is equal to $AC_{t}\times\rho_t$, where $AC_{t}$ is the accumulated capital at the end of the $(t-1)$-th holding period. More details about the action execution are provided in Section \ref{subsec:procedure}.

    \item Transition function $P$: Note that a state consists of three parts: $\bm{x}^s$, $\bm{x}^m$, and $\bm{x}^a$. 
    Since we assume that the agent's behaviours could hardly influence the market,
    the transitions of asset features $\bm{x}^s$ and market features $\bm{x}^m$ are dominated by the price movements in the market. In contrast, the transition of account feature $\bm{x}^a$ is affected by actions.
   
    \item Reward function $R$: We adopt the \textit{Differential Sharpe Ratio} ($DSR$) \cite{moody1998reinforcement} as rewards.  Recall that the \textit{Sharpe Ratio} ($SR$) is the average return in excess of the risk-free return per unit of volatility. 
    Considering an investment process that contains $T$ holding periods, its Sharpe ratio can be calculated as:
    \begin{equation}
    \label{eq: SR}
        SR_T  = (\frac{1}{T}\sum_{t=1}^{T}RoR_t  - RoR^f)/{V_T}, \\
    \end{equation}
    where $RoR_t$ is the \textit{rate of return} for the $t$-th holding period:
    \begin{equation}
        RoR_t = (AC_{t+1} - AC_{t}) / AC_{t};
    \end{equation}
    $RoR^f$ is the risk-free return rate per holding period;  and $V_t$ is the volatility:
    \begin{equation*}
    V_T = \sqrt{{\sum_{t=1}^{T}(RoR_t - \frac{1}{T}\sum_{t=1}^{T}RoR_t)^2}/{T}}.
    \end{equation*}

    Note that directly taking the Sharpe ratio as the reward induces a sparse-reward challenge for RL algorithms since it can only be obtained at the end of the trading process. Therefore, to leverage dense rewards and achieve a balance between profits and risks, we adopt the DSR as rewards. 
    Let $SR_t$ denote the Sharpe ratio at timestep $t$. The differential Sharpe ratio $DSR_t$ is then obtained by expanding $SR_t$ with decay rate $\eta$, and considerding the moving average of the return rates and the standard deviation of return rates in equation (\ref{eq: SR}),
    \begin{equation}
    DSR_t: = \frac{\mathrm{d}SR_t}{\mathrm{d}\eta} = \frac{\beta_{t-1}\Delta \alpha_t - \frac{1}{2}\alpha_{t-1}\Delta \beta_t}{(\beta_{t-1}-\alpha_{t-1}^2)^{\frac{3}{2}}},     
    \end{equation}
    where $\alpha_t$ and $\beta_t$ are exponential moving estimates of the first and second moments of $R_t$:
    $$  \alpha_t = \alpha_{t-1} + \eta \Delta \alpha_t = \alpha_{t-1} + \eta(R_t - \beta_{t-1}),$$
    $$  \beta_t = \beta_{t-1} + \eta \Delta \beta_t = \beta_{t-1} + \eta(R_t^2 -\beta_{t-1}).$$
    With attractive properties such as weights recent returns more and facilitating recursive updating, the DSR is a proven effective reward form as it enables efficient online optimization \cite{moody1998reinforcement}.

\end{itemize}

In the above MDP environment, after taking action $a_t$ at state $s_t$, a trading agent switches to the next state $s_{t+1}$ according to the transition function $P(s_{t+1}|s_t, a_t)$ and obtains a reward $r_t=R(s_t, a_t, s_{t+1})$.
To solve the portfolio management problem, we need to optimize a policy $\pi(a_t|s_t)$ which outputs an action $a_t$ for a given state $s_t$. 
To balance profits and risks, we adopt the differential Sharpe ratio as rewards. The learning objective is to obtain a policy $\pi(a_t|s_t)$ to maximize the cumulative rewards $\mathbb E_{\pi}[\sum_{t=0}^{T-1}\gamma^t r_t]$, where $T\times\tau$ is the ending time of the whole trading process.

\subsection{Trading Procedure}
\label{subsec:procedure}
Consider the investment process on a portfolio that contains both long position and short position.  
At the end of the $(t -1)$-th holding period, an investor holds ${\bf b}_{t-1}^+ = \{b_{t-1,1}^+, b_{t-1,2}^+, ...,$ $ b_{t-1,n}^+\} \in \mathbb R^n$  volume of stocks on a long position. Besides, the investor has borrowed ${\bf b}_{t-1}^-  \in \mathbb R^n$ volume of stocks at timestep $t-1$ and sold them as a short position. 
Given the close price of assets $P^{close}_{t} \in \mathbb R^n$ and portfolio vector $[\bm{w}_t^+;\bm{w}_t^-]$, the investor follows the following steps to conduct wealth reposition:
(1) At the end of the $(t-1)$-th holding period, the investor sells all long position (${\bf b}_{t-1}^+$) and gets cash;
(2) Meanwhile, he/she buys the borrowed stocks (${\bf b}_{t-1}^-$) and returns them to the stock brokerage, and at that moment the wealth is denoted by $AC_{t}$.
(3) At the beginning of the $t$-th period, the investor borrows stocks from a broker according to short proportion $\bm{w}_t^-$ and wealth $AC_{t}$, and sell them immediately to get cash. At that moment the total value of the cash is denoted by $TC_t$;
(4) Afterwards, he/she purchases stocks using the cash $TC_t$ according to long proportion $\bm{w}_t^+$.

\begin{figure*}[t!]
  \centering
  \includegraphics[width=0.9\linewidth]{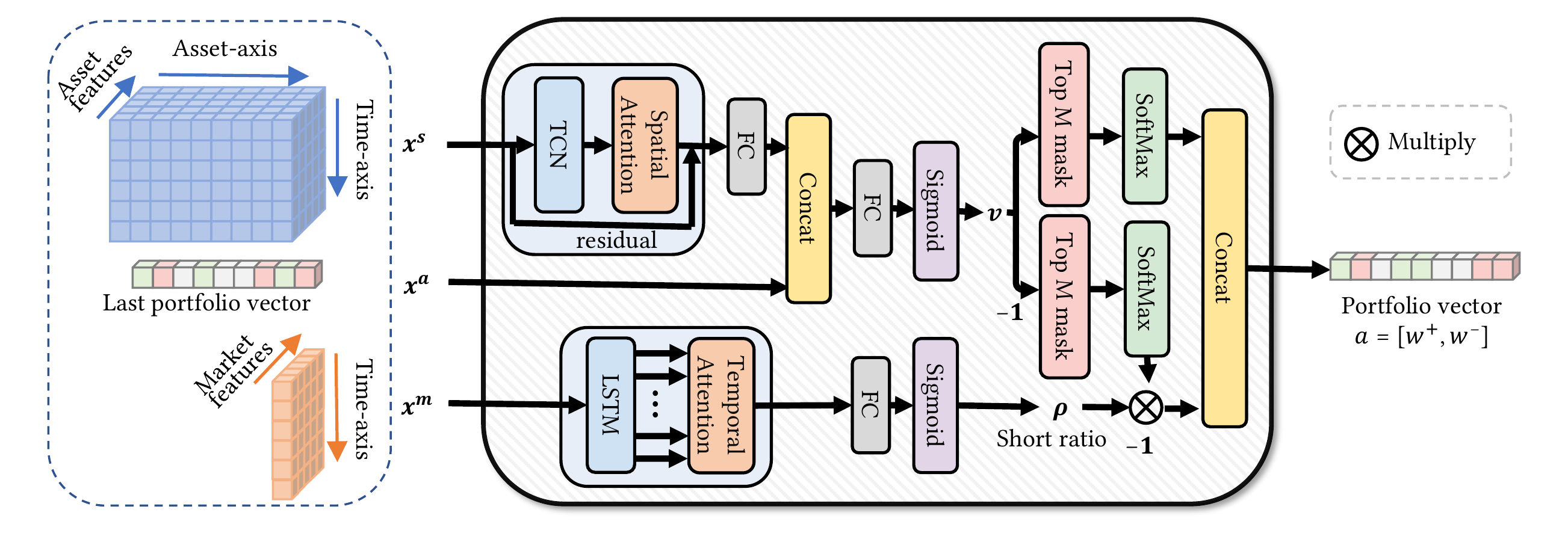}
  \setlength{\belowcaptionskip}{-0.3cm} 
  \caption{The proposed network structure for portfolio generation in the diverse policy learning phase.}
  \label{fig2}
\end{figure*}

\section{Approach}
\label{sec3}

Due to the complexity and volatility of trading markets, portfolio management has long been a challenging problem \cite{moody1998reinforcement, wang2021deeptrader, wang2019alphastock}.
A portfolio manager who follows a fixed trading strategy all the time can hardly always gain profits in the changing markets \cite{lee2020maps}. In contrast, timely identifying market conditions and switching to appropriate trading strategies potentially enable more competitive trading decisions. 
Inspired by this, we propose a novel RL-based portfolio management approach named MetaTrader, which conducts the learning by two phases, i.e, a \textit{diverse policy learning} phase and a \textit{meta-policy learning} phase.
In the first phase, MetaTrader learns multiple diverse trading policies via combining a reinforcement learning (RL) objective and an imitation learning objective. 
In the second phase, MetaTrader learns a meta-policy to select suitable policies to execute from the learned trading policy set, and the meta-policy is conditioned on the current market condition.
Figure \ref{fig1} depicts the learning scheme of those two stages.

In the rest of this section, we provide the details of the proposed approach.
Section \ref{imi} elaborates on the method of learning multiple policies which act as diversely as possible, and Section \ref{ensemble} presents the meta-policy learning, which is in charge of selecting proper policies to execute based on the market conditions.

\subsection{Diverse Policy Learning}
\label{imi}

To steadily gain profits under different market conditions, we need multiple trading policies with a great diversity. 
In general, investors face two major challenges to achieve this goal: (1) How to diversify the learned policies while maintaining their profitability; (2) How to model the asset relations over time accurately.

\subsubsection{Learning objective}
\label{obj1}
To address \textit{Challenge (1)}, we develop a novel learning objective, which combines an imitiation learning objective that encourages mimicking the behaviors in expert datasets,  with an RL objective of maximizing the cumulative rewards. The data flow of the diverse policy learning phase is shown in Figure \ref{fig1}(a). 
Here we explain the motivation of this proposed learning objective. On the one hand, despite the RL objective could help balance profits and risks, the full exploration in the complex trading environment is challenging for RL algorithms.
 Furthermore, learning via optimizing a single RL objective could not achieve the goal of generating multiple policies that act diversely. On the other hand, extracting trading knowledge from various expert datasets might help diversify the learned policies. 
 
 To acquire diverse and profitable trading policies, MetaTrader integrates the knowledge from a set of expert datasets $\{D_k|1\leq k\leq K\}$ with the policy learned online by RL. The various datasets ensure the diversity of the learned policies, and the RL objective enables the maximization of profits while balancing risks. 
 
 Sepcifically, the loss function for this mixed learning objective is formulated as
\begin{equation}
\begin{aligned}
    L_{\theta_k} = &\mathbb E_{\tau \sim \pi}[-\sum_{t=0}^T \Psi_t log\pi(a_t|s_t;\theta_k)] \\
    & + \lambda \mathbb E_{(s_i, a_i) \sim D_k} [(\pi(a|s_i; \theta_k) - a_i)^2],
    \label{eq1}
\end{aligned}
\end{equation}
 where $ \Psi_t = \sum_{t'=t}^T \gamma^{t'-t} r_{t'}$ is the discounted return,
 $\tau=(s_0, a_0, ..., s_{T+1})$ is a trajectory rolled out by policy $\pi(a_t|s_t;\theta_k)$ in the trading environment, $D_k$ is the $k$-th expert dataset and $\lambda$ is a scaling factor. 
 
The first term in Equation \eqref{eq1} is the loss function for the deterministic policy gradient algorithm \cite{sutton2018reinforcement}, which serves to maximize the cumulative rewards through optimizing policy $\pi$. The second term in Equation \eqref{eq1} is the loss for \textit{behavior cloning} \cite{pomerleau1991efficient}, which aims to extract trading knowledge from expert datasets. By training with $K$ different expert datasets, our approach obtains a policy set $\{\pi(a|s;\theta_{k})|1\leq k \leq K\}$ containing $K$ diverse trading policies, as illustrated in Figure \ref{fig1}(a). Here Policy $\pi(a|s; \theta_k)$ denotes the policy trained by imitating the behaviors in dataset $D_k$.

\subsubsection{Network Structure}
To tackle \textit{Challenge (2)}, we provide a well-designed neural network structure to abstract the time-axis and data-axis correlations among the assets. This structure is shown in Figure \ref{fig2} as a detailed supplement for Figure \ref{fig1}(a).

\textbf{Temporal Convolution Block.}
As demonstrated in Figure \ref{fig2}, MetaTrader first utilizes a temporal convolution network (TCN) \cite{yu2015multi} block to extract the time-axis relations in the \textit {asset features} data $\bm{x}^s$. Compared to recurrent neural networks, TCN has several appealing properties including facilitating parallel computation, alleviating the gradient exploration/vanishing issue, and demonstrating longer effective memory \cite{yu2015multi,wang2021deeptrader}. 
After conducting TCN operations on $x^s$ along the time axis, we obtain an output tensor denoted by $\hat{\bm H} \in \mathbb R^{N\times F \times T}$, where $F$ is the dimension of hidden features, $N$ is the number of risky assets, and $T$ is the temporal dimension.  

\textbf{Spatial Attention Layer.} 
Afterward, MetaTrader adopts an attention mechanism \cite{vaswani2017attention} to handle the spatial relationships among different assets.  Given the output vector of TCN, we calculate the spatial attention weight as 
\begin{equation}
    \hat{\bm{S}} = \bm V_s \cdot \text{sigmoid}\big((\hat{\bm H}\bm W_1)\bm W_2(\bm W_3\hat{\bm H}^{T_{1,2}})^T + \bm b_s \big) ,
\end{equation}
where $\bm W_1 \in \mathbb R^{T}$, $\bm W_2 \in \mathbb R^{F\times T}$, $\bm W_3 \in \mathbb R^{F}$, and $\bm V_s \in \mathbb R^{N\times N}$ are parameters to learn, $\bm b_s \in \mathbb R^{N\times N}$ is the bias vector, and the superscript $^{T_{1,2}}$ denotes transposing the first two dimensions of $\bm H$. The matrix $\bm \hat{S} \in \mathbb R^{N\times N}$ is then normalized by rows to represent the correlation among stocks:
\begin{equation}
    \bm S_{i,j} = \frac{\exp(\hat{\bm S}_{i,j})}{\sum_{u = 1}^N \exp(\hat{\bm S}_{i,u})}, \forall 1\leq i \leq N.
\end{equation}

\textbf{Residual Connections.} We adopt the ResNet \cite{he2016resnet} structure to alleviate the vanishing gradient problem in deep learning. To be precise, the final representation abstracted from $x^s$ is denoted by $H = S\times \hat{\bm H} + x^s$, and it is then translated to a vector with dim N using a fully connected layer:
$\hat{\bm v} = W_4\bm H + \bm b_4.$
The representation $\hat{\bm v}$ is concatenated with the \textit{account feature} $\bm{x}^a$  to obtain a vector $\bm{v}$ with dim N: $\bm v = \text{sigmoid}(W_t \cdot [\hat{\bm v};x^a] + \bm b )$, which implies the price rising potential of the N risky assets. 

Next, MetaTrader generates the portfolio weights according to vector $\bm v$: Taking the top $M$ elements of $\bm{v}$ and normalizing them with a SoftMax activation function, we obtain the weights $\bm{w}^+$ for long position; Note that $\bm v \in [-1,1]^N$, the weights $\bm{w}^-$ are acquired by a similar procedure using SoftMax activation for  $-\bm v$, and these elements are multiplied by  the total short ratio $\rho$. Let $\mathcal G^+$ and $\mathcal G^-$ denote the set of selected stocks for long position and short position respectively.  Thus, the portfolio weights can be calculated by
\begin{equation}
    w_i^+ \!=\!  \left\{ 
        \begin{array}{ll}
            \frac{\exp(\bm v_i)}{\sum_{j\in {\mathcal G^+} } \exp(\bm v_j)} & i\in \mathcal G^+ \\
          0 & i\not\in \mathcal G^+\\
        \end{array}
    \right.
    \\
     w_i^- \!= \! \left\{
        \begin{array}{ll}
            \rho * \frac{\exp(-\bm v_i)}{\sum_{j\in {\mathcal G^-} } \exp(-\bm v_j)} & i\in \mathcal G^- \\
          0 & i\not\in \mathcal G^-
        \end{array}
    \right.
\end{equation}

As shown in Figure \ref{fig2}, we condition the short ratio $\rho$ on \textit{market features} $\bm{x}^m$. Similar to DeepTrader \cite{wang2021deeptrader} and DA-RNN \cite{qin2017darnn}, MetaTrader employs an LSTM neural network \cite{hochreiter1997long} and a temporal attention structure to extract representation for market features:
\begin{equation}
 \begin{aligned}
     &\bm h_t = \text{LSTM}(\bm h_{t-1},x^m_{t}), 1\leq t \leq T, \\
    &e_t = \bm V_e^T \tanh(\bm W_5[\bm h_t;\bm h_T]) + \bm W_6x^m_k,\\
    &\alpha_t = \frac{\exp(e_t)}{\sum_{j=1}^T\exp(e_j)},\\
    &\rho_t = \frac{1}{2}\text{sigmoid}(W_7\cdot (\sum_{t=1}^T \alpha_t\cdot \bm h_t) + \bm b_m) + \frac{1}{2},\\
 \end{aligned}
\end{equation}
where $\bm V_e \in \mathbb R^{F}$,  $\bm W_5 \in \mathbb R^{F\times 2F}$, $\bm W_6 \in \mathbb R^{F\times 2F}$, $\bm W_7 \in \mathbb R^{F}$ are free parameters. 
$\bm h_t$ denotes the hidden embedding encoded by LSTM at step $t$, $\alpha_t$ is the normalized attention weight output by the temporal attention module, and $\rho_t$ is the short ratio at step $t$.
 It is worth mentioning that compared to the DeepTrader method, MetaTrader does not rely on a separate portfolio generator module. 
 The short ratio $\rho$ serves as a latent variable in the neural network, which is trained jointly with the portfolio vector $[\bm{w}^+, \bm{w}^-]$ with the learning objective introduced in Section \ref{obj1}.

\subsection{Meta-Policy Learning}
\label{ensemble}

With the diverse policies learned in Section \ref{imi}, there are generally two methods for policy ensemble: weighted sum of the pretrained policies, or selecting one policy to follow per timestep.
We adopt the latter one in this paper, as illustrated in Figure \ref{fig1}(b).

Dynamical policy selection is also a challenging sequential decision making problem. In the second phase, DeepTrader trains a \textit{meta-policy} to accomplish the automatic policy selection. The meta-policy makes decisions in a meta MDP environment. 
The meta-policy recognizes the market conditions mainly based on several macro indicators such as the index trend and market sentiment, and the historical performance of base policies. Therefore, the state at timestep $t$ in the meta-MDP is defined by $s_t^m = [\bm{x}^m_t;\bm{x}^p_t]$, where the superscript $p$ denotes the past performance of the frozen diverse policies.
And the action space for the meta-MDP is the policy identity space $\{1, ..., K\}$. Since the action space of the meta-policy is discrete, we optimize the meta-policy with the deep Q-learning algorithm \cite{mnih2015human}.
For the Q network structure, we adopt an LSTM network and the temporal attention mechanism to extract a latent embedding from the market features $\bm{x}^m_t$. The latent embedding of $\bm{x}^m_t$ is concatenated with the past performance vector $\bm{x}^p_t$. The concatenated embedding is passed through two fully connected layers to output the Q values.

\begin{algorithm}[htbp]
 	\caption{Meta-Policy Learning} \label{algo}
 	\begin{flushleft}
 	  \textbf{Input:} A trading policy set $\{\pi(a|s; \theta_k), k=1, ..., K \}$  \\
  \textbf{Initialize:} Q-function $Q(s^m, k)$, replay buffer $B$
  \end{flushleft}
 	\begin{algorithmic}[1]
 		\FOR{$episode=1..N_e$}
 		\FOR{$t=1..T$}
 		\STATE{With a probability $\epsilon$, randomly sample $k_t$ from $\{1, ..., K\}$}
 		\STATE{otherwise $k_t=argmax_k Q(s_t^m, k)$}
 		\STATE{$a_t \sim \pi(\cdot|s_t;\theta_{k_t})$}
 		\STATE{Execute the action $a_t$ and observe reward $r_t$ and new states $s_{t+1}^m$ and $s_{t+1}$}
 		\STATE{Store transition $(s_t^m, k_t, r_t, s_{t+1}^m)$ in $B$}
 		\STATE{Perform one step of optimization using Eq. \eqref{q}}
 		\ENDFOR
 		\ENDFOR
 		\STATE{\textbf{Return:} $Q(s^m, k)$}
 	\end{algorithmic}
 \end{algorithm} 

Algorithm \ref{algo} demonstrates the learning process of the meta-policy. To explore the diverse policies, MetaTrader randomly selects a policy in the policy set probabilistically (Line 3). As the Q-learning converges, the probability $\epsilon$ of the uniform random exploration decreases. 
After making decisions using the meta-policy, the portfolio weights given by the selected policy are executed in the simulated trading environment, and the agent receives a reward and transits to the next state (Line 5, 6). The transitions are stored in a replay buffer $B$ (Line 7). Q-values of the meta-policy are updated using the minibatches sampled from $B$ with the following loss function:
\begin{equation}
\label{q}
    L_Q = \mathbb{E}_{(s_j, k_j, r_j, s_{j+1}) \sim B}[(y_j - Q(s_j^m, k_j))^2],
\end{equation}
where $y_j = r_j +\gamma \max_{k_{j+1}}\hat{Q}_{target}(s_{j+1}^m, k_{j+1})$. The target Q-function $Q_{target}$ is synchronized from the optimized Q-function every $C$ steps. As $Q(s^m, k)$ is the expected future cumulative reward of selecting the $k$-th policy at state $s^m$, the learned meta-policy $\pi^m(k|s^m)$ is determined by $argmax_k Q(s^m, k)$. The portfolio weights are generated with a  joint inference of  the Q-network and the selected policy network. In addition, to stabilize the learning process, those diverse trading policies are not finetuned in this learning stage.

\section{Experiments}
\label{exp}

In this section, we conduct a series of experiments to evaluate the proposed approach. First, we demonstrate that our approach achieves better risk-return balancing in three real financial markets and substantially outperforms the baseline methods. Next, we dig into the reason for the effectiveness of the proposed method and analyze the learned portfolio management strategy in detail. Finally, we conduct ablation studies and show the importance of the meta-policy learning.

\subsection{Experimental Setup}
\label{exp_setup}

This subsection provides a description about the market data for training and test, the expert datasets for imitation learning, the compared baseline methods and the evaluation metrics.

{\bf{Financial Data. }}
Aiming at enhancing the indexes as lots of real-world funds, we conduct experiments on the constituent stocks of three well-known stock indexes, including Dow Jones Industrial Average (DJIA) in the U.S. market, Hang Seng Index (HSI) in Hong Kong market, and CSI100 Index in Chinese A-share market. 
Note that Chinese A-share market does not allow short positions, the short ratio $\rho$ is set as $0$ in this scenario.
The splitting of the data to training and test sets is shown in Table \ref{tab:dataset}.
A few companies are dropped from the pools for the sake of 
missing data \footnote{7, 12, 18 stocks are dropped from the index constitutes of DJIA, HSI, and CSI100, respectively. }. 

We adopt several technical indicators to generate input features from the price-volume data, such as MACD and KDJ, using StockStats\footnote{More indicators can be find in \url{https://github.com/dventimiglia/StockStats}.}.
\begin{table}[htbp]
  \caption{Training/Test splitting of the datasets}
  \label{tab:dataset}
  \resizebox{0.4\textwidth}{!} 
  { 
  \begin{tabular}{cccc}
    \toprule
    Index  &  \# stocks & Training & Test \\
    \midrule

    DJIA     & 23 & 1987.08-2007.08  & 2007.09-2018.12 \\
    HSI     & 38 & 2005.06-2015.04  & 2015.05-2021.12 \\
    CSI100   & 82 & 2006.01-2013.12  & 2014.01-2020.12 \\
    
    \bottomrule
  \end{tabular}
  }
\end{table}



\begin{table*}[t!]
  \caption{The comparison results on DJIA, HSI, and CSI100.}
  \label{tab:summary}
  \resizebox{\textwidth}{!} 
  { 
    \begin{tabular}{l|rrrrrr|rrrrrr|rrrrrr}
    
    \toprule
    Data & \multicolumn{6}{c|}{DJIA} & \multicolumn{6}{c|}{HSI} & \multicolumn{6}{c}{CSI100} \\
    
    \hline
     Methods & ARR(\%) & {\ul AVol} & ASR & SoR & {\ul MDD(\%)} & CR & ARR(\%) & {\ul AVol} & ASR & SoR & {\ul MDD(\%)} & CR & ARR(\%) & {\ul AVol} & ASR & SoR & {\ul MDD(\%)} & CR \\
    
    \midrule
    Market & 8.66 & 0.176 & 0.491 & 7.560 & 53.78 & 0.161 & -1.19 & 0.190 & -0.063 & -0.934 & 35.15 & -0.034 & 16.80 & \textbf{0.234} & 0.717 & \textbf{10.966} & 43.75 & 0.384 \\
    
    CSM & 12.84 & 0.537 & 0.239 & 2.073 & 55.50 & 0.231 & 4.41 & 0.391 & 0.113 & 0.426 & 40.52 & 0.109 & 32.49 & 0.510 & 0.638 & 2.945 & 66.78 & 0.487 \\
    
    BLSW & 10.78 & 0.239 & 0.452 & 1.757 & 49.73 & 0.217 & -0.35 & 0.345 & -0.010 & -0.035 & 67.01 & -0.005 & 11.21 & 0.416 & 0.270 & 1.002 & 75.01 & 0.149 \\
    
    \hline
    
    LightGBM & 8.05 & 0.251 & 0.320 & 1.051 & 62.96 & 0.128 & 5.64 & 0.313 & 0.180 & 0.611 & 48.20 & 0.117 & 14.48 & 0.464 & 0.312 & 1.103 & 72.99 & 0.198 \\
    
    DA-RNN & 12.09 & 0.230 & 0.525 & 1.841 & 47.58 & 0.254 & -6.96 & 0.281 & -0.248 & -0.723 & 60.48 & -0.115 & 13.93 & 0.305 & 0.456 & 1.630 & 42.92 & 0.325 \\
    
    \hline
    AlphaStock & 17.86 & 0.150 & 1.190 & 4.333 & 24.51 & 0.729 & 18.75 & 0.193 & 0.969 & 4.169 & 28.90 & 0.649 & 27.51 & 0.250 & 1.099 & 4.501 & \textbf{19.08} & 1.442 \\
    
    DeepTrader & 14.90 & \textbf{0.122} & 1.122 & 4.617 & \textbf{13.22} & 1.127 & 21.68 & \textbf{0.189} & 1.145 & 4.322 & \textbf{23.24} & 0.933 & 33.55 & 0.280 & 1.197 & 4.380 & 22.70 & 1.478 \\
    
    \textbf{Ours} & \textbf{25.61} & 0.196 & \textbf{1.310} & \textbf{5.602} & 19.91 & \textbf{1.287} & \textbf{44.12} & 0.320 & \textbf{1.379} & \textbf{7.759} & 27.46 & \textbf{1.607} & \textbf{43.21} & 0.249 & \textbf{1.733} & 6.409 & 22.45 & \textbf{1.924}\\
    \bottomrule
    \end{tabular}
  }
\end{table*}

{\bf{Expert Datasets. }}
To acquire diverse trading policies, we employ four different datasets to train our approach.
Three expert datasets ($D_1\sim D_3$)  are used to generate policies imitating the momentum, mean reversion and hindsight strategies respectively. 
An empty dataset $D_4$ is also adopted to generate a policy learned from interactions with the environment without imitation. 

\begin{itemize}
    \item ${\bf D_1}$ is a set of portfolio vectors following the Cross-Sectional Momentum (CSM) \cite{jegadeesh2002csm} strategy during the training period. 
    In the CSM method, stocks are ranked based on their relative strength momentum, and we utilize the past price rising rate $\prod \limits_{i=1}^3(1 +RoR_{t-i})$ as the momentum when make decisions at timesetp $t$.
    The long/short portfolios consist of the top/bottom $M$ stocks with equal weights.

    \item ${\bf D_2}$ is a set of portfolio vectors following the Buying-Loser-Selling-Winner (BLSW) \cite{jegadeesh1993blsw} strategy during the training period. BLSW is a classical mean reversion strategy.
    Stocks are ranked according to the difference between prices and their near-term averages. The long/short portfolios consist of the top/bottom $M$ stocks with equal weights.
    
    \item ${\bf D_3}$ is a set of portfolio vectors decided by a hindsight greedy strategy. As we are accessible to the price data during the training time,
     we create a hindsight expert that purchases $M$ stocks with the highest future returns and sell those with the lowest future returns. The weights of the $M$ stocks are generated with the normalization of the the price rising rates, different from the equal weights in $D1$ and $D2$.
     
    \item ${\bf D_4}$ is an empty dataset. It can be considered as a special case in Equation \eqref{eq1}, i.e., $\lambda=0$.
\end{itemize}

{\bf{Baseline Methods. }}
We compare our method with the three categories of portfolio management methods summarized in Section \ref{related}. (1) {Traditional investment strategies}: \textit{Market}, CSM and BLSW. {Market} is  a strategy replicating the index. CSM and BLSW are the strategies to generate the expert datasets. (2) {Supervised learning approaches}\footnote{The implementation of those supervised learning methods is based on Qlib (\url{https://github.com/microsoft/qlib}).}: LightGBM \cite{ke2017lightgbm} and DA-RNN \cite{qin2017darnn}.
LightGBM is a widely-used ensemble model of decision trees for classification and regression. The DA-RNN method utilizes an LSTM neural network with temporal attention mechanism for stock price prediction. 
(3) {State-of-the-art RL-based methods}: AlphaStock \cite{wang2019alphastock} and DeepTrader \cite{wang2021deeptrader}.

{\bf{Evaluation Metrics. }} 
We adopt six evaluation metrics to meet different risk appetites of the investors.
\begin{enumerate}
    \item \textit{Annualized Rate of Return (ARR)} is the annualized average of the return rate of one holding period, calculated as 
    $$ARR_T = (\overline{RoR}_{1:T} - RoR^f)\times N_y,$$
    where $\overline{RoR}_{1:T}$ is the average return rate of a holding period, and $N_y$ is the number of holding periods in a year.
    
    \item  \textit{Annualized Volatility (AVol)} is an annualized average of volatility ($Vol$) that reflects risks of a strategy
    $$AVol_T = V_T \times \sqrt{N_y}.$$ 
    
    \item \textit{Maximum DrawDown (MDD)} measures the loss under the worst case during the investment. It can be defined as
    $$MDD_T = \max_{1\leq i< j\leq T}(AC_i - AC_j)/{AC_i},$$
    where $AC_i$ and $AC_j$ denote the accumulated capital at timestep $i$ and $j$ respectively.
    
    \item  \textit{Annualized Sharpe Ratio (ASR)} is the risk-adjusted ARR based on Annualized Volatility, calculated as 
    $$ASR_T = \frac{ARR}{AVol}.$$
    
    \item \textit{Calmar Ratio (CR)} is the risk-adjusted ARR based on MDD, formulated by $$CR_T = APR/MDD.$$ 

    \item \textit{Sortino Ratio (SoR)} is also a risk-adjustment metric, which implies the additional return for each unit of downside risk, defined as 
    $$SoR =\frac{ARR_T}{ \sqrt{{\sum_{t=1}^{T}(\min(RoR_t,0) - \frac{1}{T}\sum_{t=1}^{T}\min(RoR_t,0))^2}/{T}}}.$$
\end{enumerate}

Among these metrics, ARR is a profit criteria; AVol and MDD are risk criterion for which the lower values are preferred; ASR, CR, and SoR are risk-return criterion.

\subsection{Comparison Results on DJIA, HSI, and CSI100}
\label{res}

\begin{figure*}[h!]
  \centering
  \subfloat[DJIA.]{\label{fig:3a}\includegraphics[scale=0.3]{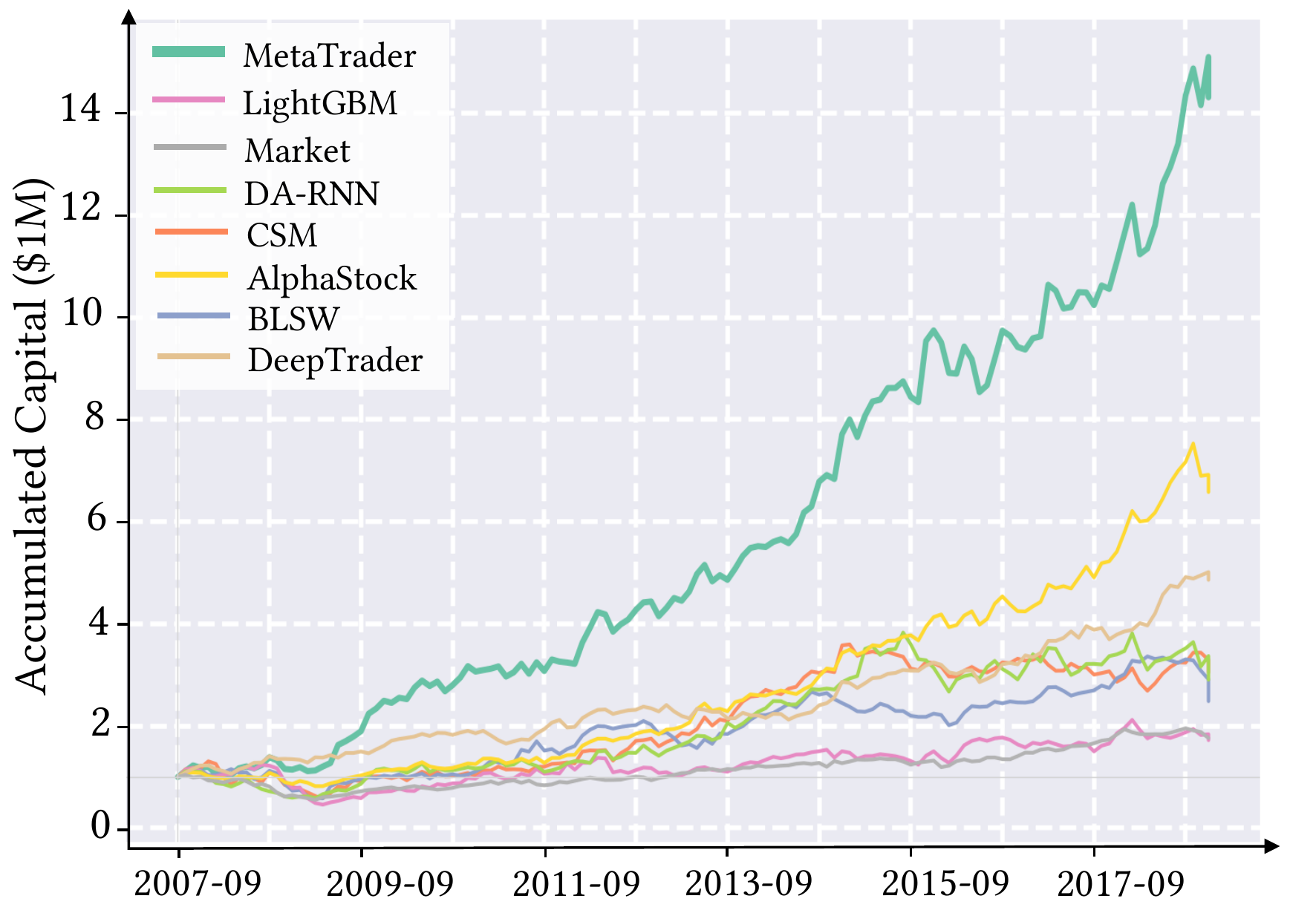}}
  \subfloat[HSI.]{\label{fig:3b}\includegraphics[scale=0.3]{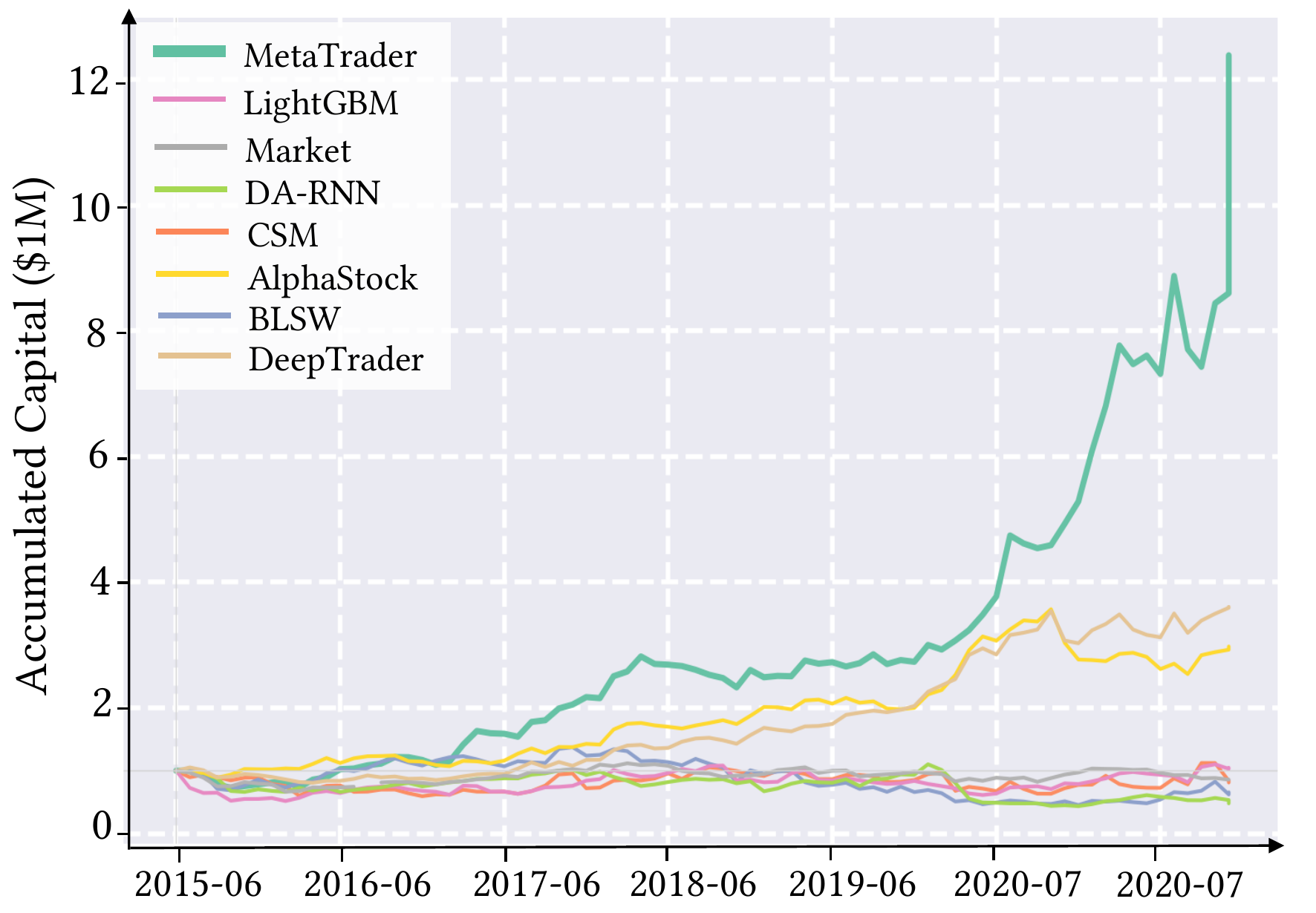}}
  \subfloat[CSI100.]{\label{fig:3c}\includegraphics[scale=0.3]{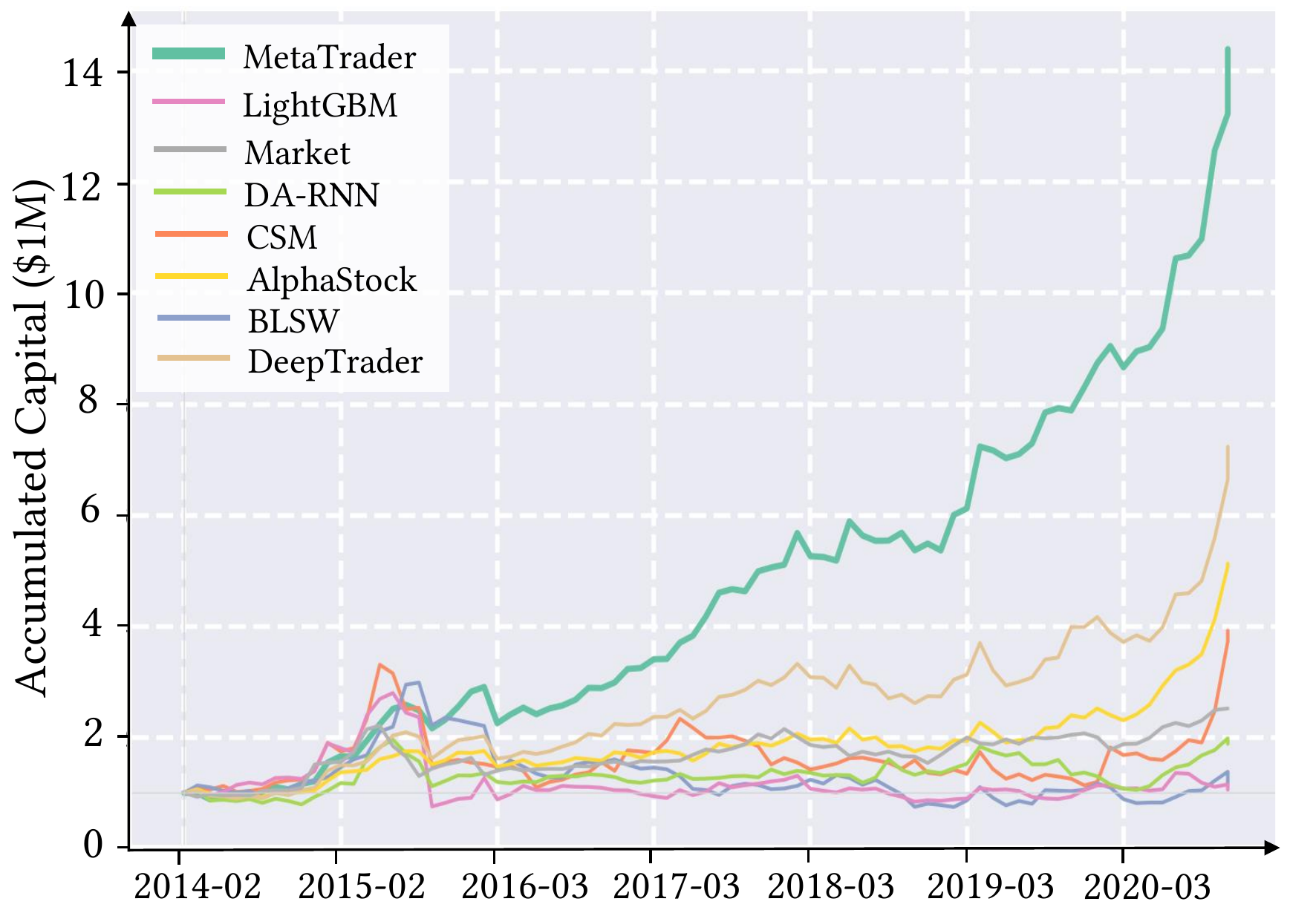}}
  \caption{The accumulated capital of the proposed method and baselines on DJIA,HSI and CSI100.}
  \label{fig:comparison}
\end{figure*}

Table \ref{tab:summary} summarises the comparison results of our approach and the baselines, and Figure \ref{fig:comparison} depicts the corresponding accumulated capital in the three markets.

{\bf Performance on DJIA. }  
Our method achieves the best performance in terms of profit criteria (ARR) and profit-risk criteria (ASR, SoR, and CR). It gains a much higher return rate while keeping the risk at a relatively lower level compared to other methods. For the traditional investment strategies, the CSM strategy achieves a higher average return rate than the Market. However, it also suffers from high volatility and MDD.  
As shown in Figure \ref{fig:comparison}(a), the CSM strategy performs well from  2009 to 2014, but it performs poorly from 2015 to 2018.    While the BLSW strategy performs well during the period of 2009-2012, 2013-2014, and 2015-2017, and fails to gain profits in 2012 and 2014. These results verify that the traditional investment strategies only perform well in certain cases. 
The supervised learning-based baselines (LightGBM and DA-RNN) fail in reducing risks, since they concentrate more on price movement prediction, ignoring the expensive cost of failed predictions. 
We find that RL-based methods, i.e., AlphaStock, DeepTrader, and our method, outperform traditional strategies and supervised learning-based methods in risk-return balancing. In Figure \ref{fig:comparison}(a), the wealth curves of RL-based methods rise steadily.  Correspondingly, as shown in the DJIA columns of Table \ref{tab:summary}, these methods have high risk-adjusted return rates and low risks. In particular, DeepTrader achieves the lowest MDD ($13.22\%$) possibly because MDD is adopted as the reward in its market scoring unit. 
This phenomenon validates the flexibility and effectiveness of the RL objective in portfolio management.
Since our method could extract knowledge from the datasets and tackle different market conditions with the corresponding trading styles,
 our method has accomplished a significantly larger ARR than DeepTrader at the cost of slightly greater risks, leading to higher profit-risk criteria.

{\bf Performance on HSI. } As shown in the middle columns of Table \ref{tab:summary}, the ARR of {Market} on HSI in the back-test period is negative, which indicates a grim economic situation. Even in this situation, our method still substantially outperforms the baselines in terms of profits and profit-risk criteria, verifying the effectiveness of our approach. As illustrated in Figure \ref{fig:comparison}(b), the traditional investment strategies and the supervised learning baselines perform no better than the Market. The two RL baselines remain competitive methods in this bad economic situation.

{\bf Performance on CSI100. } In the market of CSI100, CSM achieves a fascinating ARR of 32.49\%. However, as demonstrated in the right columns of Table \ref{tab:summary}, all the traditional methods and supervised learning methods have poor performance regarding MDD and AVol. These results are consistent with the declining wealth curves during the bear market in 2015 shown by Figure \ref{fig:comparison}(c).
 In comparison, RL-based methods maintain relatively steady performance during the whole back-test period. Particularly, our method outperforms the baselines in obtaining higher profits with low risks. 

\vspace{-1cm}

\subsection{Analysis of Diverse Trading policies}
\label{div}
We visualize the trading policies trained with the learning objective proposed in Section \ref{imi} to answer the following questions: {\bf (1)} Do the learned policies have diverse trading styles? {\bf (2)} Is there a single policy performing consistently better than others from the beginning to the end?

\begin{figure}[htbp]
  \centering

  \subfloat[Portfolio vectors given by the four trading policies on May 18th, 2010.]{\label{fig:4a}\includegraphics[scale=0.23]{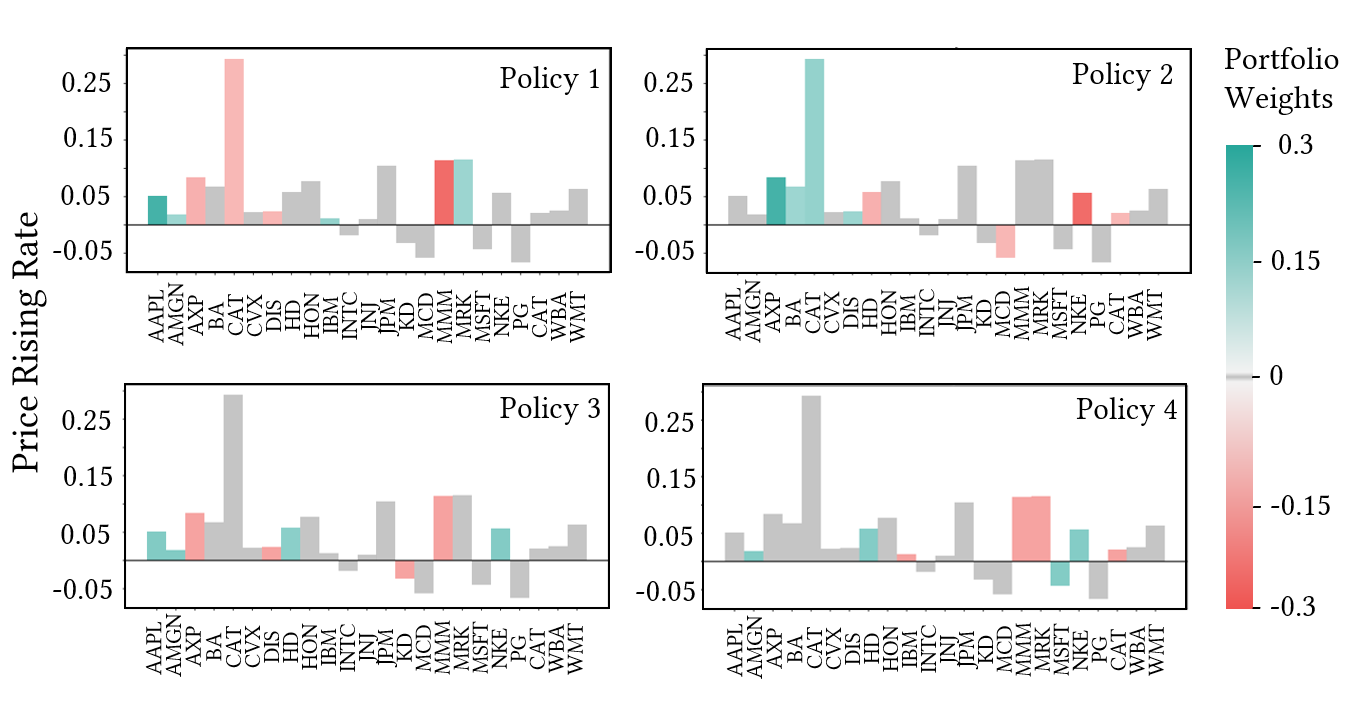}}\\
  \subfloat[The monthly return rates of the learned base policies from Oct., 2007 to Sep.,2010.]{\label{fig:4b}\includegraphics[scale=0.26]{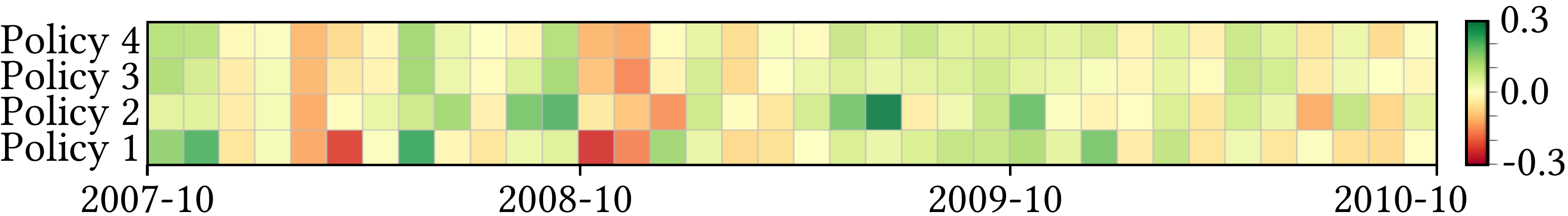}}\\
  \subfloat[Ranking the four policies by profits from Oct., 2010 to Sep.,2013. Darker colors denote higher monthly profits (larger ranks) .]{\label{fig:4c}\includegraphics[scale=0.26]{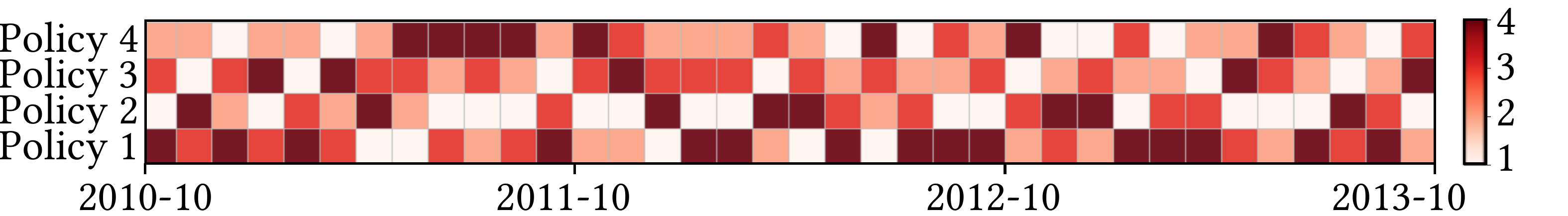}}\\
  \caption{Visualization of the  diverse policies learned in the first phase on DJIA constituent stocks.}
  \label{fig:diverse}
\end{figure}

To answer \textit{Question} (1),  we depict the portfolio weights given by the four trading policies on May 18th, 2010 in the U.S.  market in Figure \ref{fig:diverse}(a).
The heights of the  bars represent the price rising rates in  the future trading period. The colors denote the portfolio weights of the candidate stocks: green for long operations, and red for short operations.
 As shown in Figure \ref{fig:diverse}(a), the behaviors of those policies are quite different from each other. 
For example, Policy $1$ and $2$ hold opposite views on the allocation for stock AXP and CAT. According to the price rising rates, the decision of Policy $2$ is the most reasonable on that day. That is, the stocks with higher price rising rates are selected for long operations by Policy $2$.

To answer \textit{Question} (2), we rank the four policies based on their monthly profits from Oct. 2010 to Sep. 2013 in the U.S. market in Figure \ref{fig:diverse}(c), where higher ranks indicate higher profits. Figure \ref{fig:diverse}(c) demonstrates that no single policy always performs better than the others. Instead, their rankings fluctuate with time. Besides, Figure \ref{fig:diverse}(b) depicts the monthly return  of the four base policies  from Oct. 2007 to Sep. 2010. According to Figure \ref{fig:diverse}(b), in the vast majority of cases, there are at least one base policy that performs not bad even when encountering the 2008 crisis. This confirms the effectiveness of the first phase of DeepTrader.
Therefore, switching to appropriate policies according to the current market conditions is crucial to improving the performance of portfolio management as demonstrated in the next subsection.

\subsection{Visualization of the Learned Meta-Policy}
\label{meta}

\begin{figure}[htbp]
  \centering
  \includegraphics[width=.86\linewidth]{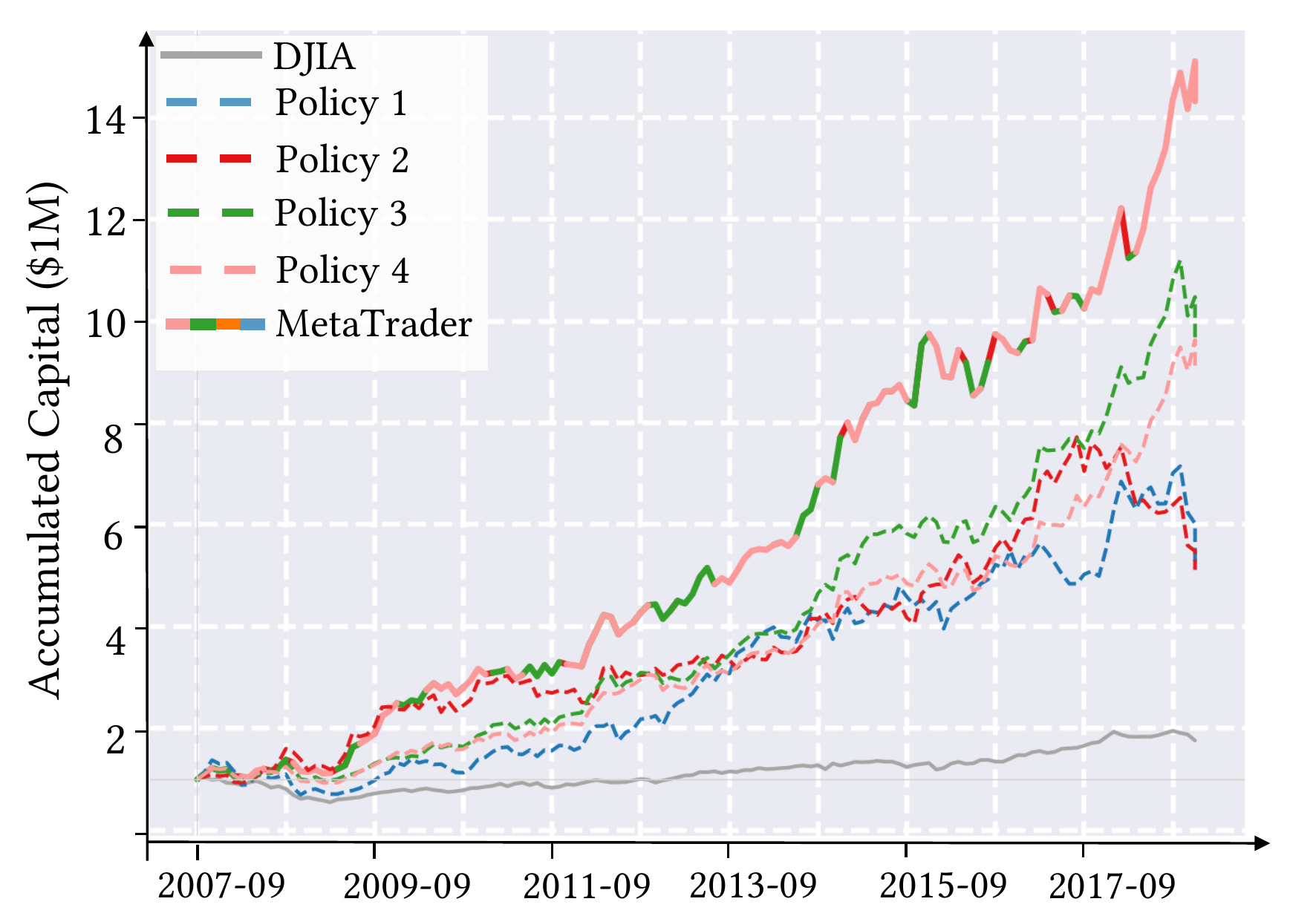}\\
  \caption{The accumulated capital of the meta-policy and the base policies on DJIA during the test period.}
  \label{fig:ensemble}
\end{figure}

In this subsection, we visualize and analyze the decisions of the learned meta-policy on DJIA during the test period.
Figure \ref{fig:ensemble} compares the accumulated capital of the four base trading policies and the method of integrating them together. The color of the integration curve represents the policy identity selected by the meta-policy.
As we can see, the wealth of the base policies all significantly outperforms that of the Market index, which verifies the utility of the novel learning objective proposed in Section \ref{imi}.
With the second training stage described in Section \ref{ensemble},
the meta-policy has learned to select the most profitable trading policy to follow. 
For example, after Sep. 2011, the policy selection switches from the red policy (Policy 2) to the lightcoral policy (Policy 4). Similarly, from Sep. 2013 to Sep. 2014, the selection of the red policy is switched back to the light blue policy for a faster increase of wealth.
We calculate the profit rankings of the selected policies as well, and find that for about $95\%$ of the trading dates, the meta-policy selects the most profitable or the second profitable policy.

\subsection{Ablation Study of Meta-Policy Learning}
\label{ablation}


\begin{table*}[t!]
  \caption{Ablation study of the meta-policy learning.}
  \label{tab:ablation}
  \resizebox{\textwidth}{!} 
  { 
    \begin{tabular}{l|rrrrrr|rrrrrr|rrrrrr}
    
    \toprule
    Data & \multicolumn{6}{c|}{DJIA} & \multicolumn{6}{c|}{HSI} & \multicolumn{6}{c}{CSI100} \\
    
    \midrule
    Methods & ARR(\%) & {\ul AVoL} & ASR & SoR & {\ul MDD(\%)} & CR & ARR(\%) & {\ul AVoL} & ASR & SoR & {\ul MDD(\%)} & CR & ARR(\%) & {\ul AVoL} & ASR & SoR & {\ul MDD(\%)} & CR \\
    
    \midrule
    Single-Best & 16.05 & \textbf{0.133} & 1.210 & 5.210 & \textbf{13.10} & 1.226 & 22.31 & 0.223 & 0.999 & 5.222 & \textbf{19.08} & 1.169 & 35.67 & \textbf{0.221} & 1.236 & 5.207 & \textbf{20.40} & 1.363 \\
    
    Random-Pick & 21.64 & 0.182 & 1.189 & 4.637 & 24.04 & 0.900 & 19.54 & 0.217 & 1.191 & 3.437 & 31.44 & 0.622 & 22.15 & 0.444 & 0.499 & 2.000 & 63.55 & 0.348 \\
    
    Average-Weight & 20.08 & 0.174 & 1.152 & 3.844 & 28.01 & 0.717 & 18.59 & \textbf{0.177} & 0.901 & 3.699 & 32.36 & 0.575 & 24.89 & 0.309 & 0.805 & 2.965 & 42.77 & 0.582 \\
    
    \textbf{Ours} & {\bf 25.61} & 0.196 & {\bf 1.310} & {\bf 5.602} & 19.91 & {\bf 1.287} & {\bf 44.12} & 0.320 & {\bf 1.379} & {\bf 7.759} & 27.46 &{\bf 1.607} & {\bf 43.21} & 0.249 & {\bf 1.733} & {\bf 6.409}& 22.45 & {\bf 1.924} \\

    \bottomrule
    \end{tabular}
  }
\end{table*}

To investigate the effectiveness of the meta-policy learning,
 we compare the proposed approach with its three variations including \textit{Single-Best}, \textit{Random-Pick} and \textit{Average-Weight}. \textit{Single-Best}  selects the policy with the largest ARR from the policy set. 
 \textit{Random-Pick} denotes using a random meta-policy instead of training it. \textit{Average-Weight} means taking the average of outputs of the four base policies as the portfolio weights, which is the same as the ensemble method in MAPS \cite{lee2020maps}.
 
 The results of the ablation study are shown in Table \ref{tab:ablation}. The performance of our method is superior than all the three variant methods. This phenomenon empirically verifies that it is critical to  effectively integrate the learned base policies based on market conditions via learning the meta-policy. Note that the \textit{Single-Best} method outperforms the other two variant methods on HSI and CSI100, indicating that integrating different policies by averaging or random selection is unhelpful to spread risks.

\section{Related Work}

\label{related}
Quantitative portfolio management has received much attention in  financial domain. We classify the related work into three categories, and provide a discussion about them as follows.

{\bf Traditional Investment Strategies.} 
Traditional portfolio management strategies comprise momentum trading, mean reversion, and multi-factor models.
The momentum trading strategies suggest following the current trends, e.g., buying the winners and selling the losers \cite{grinblatt1995momentum}, cross-sectional momentum \cite{jegadeesh2002csm}, time series momentum \cite{moskowitz2012tsm}, and so on.
The mean reversion strategy \cite{poterba1988mean} purchases assets whose prices are lower than their historical mean and sells those whose prices are higher than the mean, e.g., buying the losers and selling the winners \cite{jegadeesh1993blsw}. 
Multi-factor models \cite{Fama1996multifactor} do asset selection according to various factors correlated with the returns of the assets. 
Although based on solid financial theories, most traditional investment strategies only perform well in particular scenarios, and fail to adapt to the complex markets.

{\bf Supervised Learning in Portfolio Management.} 
With an exceptional ability to extract useful representations, deep neural networks have been applied to solve the portfolio management problem in recent years. The supervised learning methods firstly train a deep predictive model of asset prices, and then use the predictive model and a rule to generate portfolio weights. Previous supervised learning methods mainly focus on two aspects to improve the prediction accuracy: using advanced neural network structures \cite{qin2017darnn, yoo2021dtml, sawhney2020,sawhney2021hypergraph,cheng2021graphattention}, and proposing new representation learning objectives \cite{feng2018enhancing, hou2021contrastive,Trade}.

The DA-RNN method \cite{qin2017darnn} adopts a temporal attention structure to learn the correlations among historical prices, and the DTML approach \cite{yoo2021dtml} utilizes a transformer encoder to learn inter-stock correlations.
Graph neural networks \cite{scarselli2008graph} are also widely used to detect stock relations by extracting knowledge from extra data resources \cite{sawhney2020,sawhney2021hypergraph,cheng2021graphattention,hu2018news,9101395}.
As for the new representation learning objectives, \citet{feng2018enhancing} proposed an adversarial training objective to achieve more robust predictions, and \citet{hou2021contrastive} leveraged a contrastive learning objective to process the multi-granularity data in markets.
The prediction models in those supervised learning methods are trained with deep learning techniques, but the way of generating portfolio weights is based on specific rules. In contrast, our approach trains the portfolio allocation framework in an end-to-end manner, which has obtained superior performance in markets as demonstrated in Section \ref{exp}.

{\bf Reinforcement Learning in Portfolio Management.} 
Compared to the supervised learning methods, RL provides a seamless and flexible framework for portfolio management \cite{sun2021reinforcement}. 
With different risk appetite, previous RL-based portfolio management algorithms adopt various reward functions including
the Sharpe ratio \cite{wang2019alphastock}, the maximum drawdown \cite{almahdi2017adaptive, wang2021deeptrader}, and the total profits \cite{jiang2017elle, ye2020reinforcement, wang2020commission}.
The performance of the methods utilizing a pure RL objective is constrained by not fully exploring the fluctuate markets.
Via imitating the behaviors of real investors, the Investor-Imitator method \cite{ding2018ii} achieves better exploration in the complex trading environment. However, this method has not solved the problem of integrating different human trading styles  to accomplish a superior trading performance.
Besides the portfolio management methods using the single-agent RL framework, some related work employs multi-agent techniques to learn investment decisions \cite{batista2010multiagent,lee2020maps, huang2022mspm}.
MAPS \cite{lee2020maps} is a multi-agent portfolio management method, which diversifies the multiple trading agents with a global reward based on correlations between agents. Nevertheless, MAPS naively takes the average of those agents' outputs as the final portfolio weights. 
The performance of this averaging method is much worse than our method of learning a meta-policy, as demonstrated in Section \ref{ablation}.
MPSM \cite{huang2022mspm} is also a multi-agent portfolio management method, and it aims to obtain a scalable and reusable trading system, which is different from our motivation of training policies with varying trading styles to succeed in various market conditions.

\section{Conclusion}
\label{con}
In this paper, we propose MetaTrader, a two-phase deep RL-based portfolio management approach incorporated with imitation learning techniques to deal with the challenge of changing markets. 
Through abstracting trading knowledge from multiple experts meanwhile interacting with the environments, MetaTrader learns a set of base policies that act diversely in the first phase. 
Visualization results confirm the diversity of the learned base policies.   
In the second phase, these learned base policies are then integrated by a meta-policy optimized by DQN based on market conditions. 
Experiments on three markets demonstrate that MetaTrader outperforms the existing baselines in terms of risk-adjusted returns. And Further ablation studies verify the effectiveness of the components in MetaTrader. Since it is important for investors to avoid risks in financial scenarios, we expect the meta-policy could conduct policy ensemble while providing the confidence for its decisions in the future work. 
  
\section{Acknowledgement}
Niu and Li are supported in part by the National Natural Science Foundation of China Grant 62161146004, Turing AI Institute of Nanjing and Xi'an Institute for Interdisciplinary Information Core Technology.

\bibliographystyle{ACM-Reference-Format}
\bibliography{imaps}

\end{document}